\documentclass[a4paper,11pt,twoside,onecolumn]{article}

\usepackage{cite}
\usepackage{graphicx}
\usepackage{epsf}
\usepackage{amssymb}
\usepackage{color}
\usepackage{amsmath}

\onecolumn
\def\sloppy{\tolerance=100000\hfuzz=\maxdimen \vfuzz=\maxdimen}
\newcommand{\rme}{\mathrm{e}}
\newcommand{\rmi}{\mathrm{i}}
\newcommand{\rmd}{\mathrm{d}}
\newcommand{\rmD}{\mathrm{D}}
\newcommand{\Tr}{\mathrm{Tr}}
\newcommand{\bfr}{\mathbf{r}}

\newcommand{\tmu}{\widetilde{\mu}}

\vspace{-8ex}\date{}

\begin{document}
\sloppy

\title{\large\bf PARTITION FUNCTION OF THE BOSE--EINSTEIN CONDENSATE DARK MATTER AND THE MODIFIED GROSS--PITAEVSKII EQUATION}

\author{A.V. Nazarenko\\ 
{\small\it Bogolyubov Institute for Theoretical Physics of NAS of Ukraine,}\\
{\small\it 14-b, Metrolohichna str., UA--03143 Kyiv, Ukraine}\\
{\small nazarenko@bitp.kiev.ua}}

\maketitle

\begin{abstract}
Intending to describe the dark matter of dwarf galaxies, we concentrate on one model of
the slowly rotating and gravitating Bose--Einstein condensate. For a deeper understanding of
its properties, we calculate the partition function and compare the characteristics derived
from it with the results based on the solution of Gross--Pitaevskii equation. In our approach,
which uses the Green's functions of spatial evolution operators, we formulate in a unified
way the boundary conditions, important for applying the Thomas--Fermi approximation.
Taking this into account, we revise some of the results obtained earlier. We also derive
the spatial particle distribution, similar to the model with rotation, by using the deformation
of commutation relations for a macroscopic wave function and modifying the Gross--Pitaevskii
equation. It is shown that such an approach leads to the entropy inhomogeneity and makes
the distribution dependent on temperature.


\end{abstract}

\section{Introduction}

Among the famous and developed models of dark matter 
(DM)\cite{Harko19,HMKT15,KSKS15,KKPY17,RME18,A18,CM16,SNE16,SN19,Bel14,CKS16,GKV19},
we decided to focus on the Bose--Einstein condensate (BEC) DM, which is formulated on the basis
of the non-relativistic Gross--Pitaevskii equation (GPE) and, as can be seen from the recent
works~\cite{Harko11,Harko18,KKG}, well describes the observed rotation curves of dwarf galaxies.
Here we omit the history of this idea, the diverse argumentation of which has already been
given by its authors~\cite{BH07,Harko11}, as well as its direct astronomical applications.

The advantage of this model is related to the clear role of leading physical phenomena,
which are taken into account when using mathematical simplifications, which engage also
the Thomas--Fermi approximation. Basically, the behavior of the system is determined by
Newtonian gravitational attraction and repulsive scattering. Studying stationary solutions of 
the hydrodynamic type due to the Madelung representation for the condensate wave function,
the Chandrasekhar's results~\cite{Chandra33} are applied. Nevertheless, the conditions for
their application requires a detailed consideration, as a consequence of which the conclusions
are made~\cite{CH12}. First of all, the analysis made indicates the polytropic index equal to
$n=1$ for the equation of state that determines the dependence of pressure on density.
The influence of thermal fluctuations has also been established~\cite{HarMad11,HLLM15}.

Inspired by outcomes of the BEC DM model, we would like to get here a consistent description
of it, using different theoretical approaches. We consider it new to obtain the mean macroscopic
characteristics from the computed partition function, which should coincide with the results
based on the solution of GPE. However, we do not initially impose the geometric constraints
that are used in differential calculus. It is admitted that our generalization will affect
the determination of the size of boson system described in the Thomas--Fermi approximation~\cite{Harko18,BH07}.

Since the global rotation of DM has not yet received a final explanation (and is often
associated with processes at an early stage of the Universe evolution~\cite{Harko18}),
we try to reproduce the effects caused by it in a different way. More specifically,
we propose to relate these effects with the (statistical) properties of DM particles
at finite temperature. For this, we only assume that the correlation length is much larger
than interparticle distance, that is ensured by a little particle mass. Thus,
exploiting the BEC DM model, we are able to extract the parameters of DM particles from
macroscopic characteristics, if the particle distribution, taking into account a slow rotation,
describes satisfactorily the observables. In practice, we will deform the commutation relations
for the condensate wave function, that is equivalent to changing the integration measure
in the partition function. Using the connection between the two computational approaches
(based on the partition function and GPE), we intend to obtain a modified Gross--Pitaevskii
equation. In fact, temperature would replace the centrifugal energy in the expressions used.
It does not contradict the BEC concept if the temperature is still sufficiently low.
However, an entropy inhomogeneity in space is expected to be an important consequence of
such a deformation: it should reach a minimum value in regions with a high particle
density and a maximum in a rarefied medium. It is in contrast to the model with rotation,
where the entropy is zero everywhere.

On the other hand, we get the opportunity to study the consequences of deformation as
a functional of particle density given in the coordinate representation. We consider
this nontrivial because deformations in the momentum representation are usually
performed~\cite{GM15,CGN19}.

The present paper is organized as follows. In the next Section we develop the partition
function formalism for the gravitating BEC with rigid rotation and compare the macroscopic
characteristics, obtained by using the GPE solution and thermodynamics. In Section III we
deform appropriately the commutation relations for wave function and analyze the solution
of modified GPE. We end the paper with the discussion.

%%%%%%%%%%%%%%%%%%%%%%%%%%%%%%%%%%%%%%%%%%%%%%%%%%%%%%%%%%%%%%%%%%%%%%%%%%%%%%%%%%%%%%%%%%%%%
\section{The Partition Function of Gravitating Bose--Einstein Condensate}

Let us consider the partition function $\mathcal{Z}$ of the bosons with mass $m$, described
by the wave-function $\psi(\bfr)$ and slowly rotating with a cyclic frequency $\omega$
in Cartesian $(xy)$-plane at a given temperature $T=\beta^{-1}$ and chemical potential $\mu$:
\begin{eqnarray}
&&\mathcal{Z}=C\int \rmD\psi\,\rmD\psi^*J[\psi,\psi^*]\exp{\left(-\beta\, \Gamma_\omega[\psi,\psi^*]\right)},
\label{Zfirst}\\
&&\Gamma_\omega[\psi,\psi^*]\equiv E[\psi,\psi^*]-\int\nu(\bfr)\,\rho(\bfr)\rmd\bfr,\\
&&\nu(\bfr)\equiv\mu+\frac{m\omega^2}{2}(x^2+y^2),
\end{eqnarray}
where $C$ is a normalization, and the particle density
$\rho(\mathbf{r})=|\psi(\mathbf{r})|^2\geq0$ defines the total particle number:
\begin{equation}\label{Num}
\mathcal{N}[\rho]=\int\rmd\bfr\,\rho(\bfr).
\end{equation}

The Jacobian $J[\psi,\psi^*]=\mathrm{Det}\left\|\{\psi(\bfr_1),\psi^*(\bfr_2)\}\right\|$
defines geometrically the integration measure in the general case and can be expressed
in the terms of Poisson brackets~\footnote{In the classical picture these relations
replace the quantum ones, $[\hat a_\bfr,\hat a^\dag_\mathbf{s}]=\delta^3(\bfr-\mathbf{s})$,
for the creation and annihilation operators in the coordinate representation.} for
the function $\psi(\bfr)$ and its complex conjugate $\psi^*(\bfr)$:
\begin{eqnarray}
&&\{\psi(\bfr_1),\psi(\bfr_2)\}=\{\psi^*(\bfr_1),\psi^*(\bfr_2)\}=0,
\nonumber\\
&&\{\psi(\bfr_1),\psi^*(\bfr_2)\}=-\rmi\delta^3(\bfr_1-\bfr_2).
\label{PB1}
\end{eqnarray}

We examine (\ref{Zfirst}) for the boson cloud model, represented
geometrically by the ball $B=\{\bfr\in\mathbb{R}^3|\,|\bfr|\leq R\}$,
$\mathrm{Vol}(B)=4\pi R^3/3$, and given by the energy functional in
the Thomas--Fermi approximation (when the kinetic energy term is neglected):
\begin{equation}\label{EnFn}
E[\psi,\psi^*]=\frac{U}{2}\int_B\rmd\bfr\,
\rho(\bfr)\left(1+\kappa^2\Delta^{-1}\right)\rho(\bfr).
\end{equation}
Here, $U=4\pi\hbar^2a/m$ and $\kappa^2=4\pi Gm^2/U$ are the model parameters,
including the gravitational constant $G$ and the scattering length $a$;
$\Delta^{-1}$ is an inverse Laplace operator such that
$\Delta\Delta^{-1}=\Delta^{-1}\Delta=1$.

The first term of (\ref{EnFn}) represents a repulsive pair interaction,
and the second is the gravitational interaction, which can be described using
the non-relativistic gravitational potential $V(\bfr)$, which is the subject of
the Poisson equation:
\begin{equation}
\Delta V(\bfr)=4\pi G\rho(\bfr), \qquad \bfr\in B.
\end{equation}

Determining the path integral $\mathcal{Z}$, we use the Madelung transformation:
\begin{equation}\label{Mad}
\psi=\sqrt{\rho}\,\rme^{\rmi\phi},\quad
\rho\in[0;+\infty),\quad
\phi\in[0;2\pi).
\end{equation}
Since the measure
$J\,\rmD\psi\,\rmD\psi^*\sim\prod_{\{\bfr\in B\}}\rmd\rho(\bfr)\,\rmd\phi(\bfr)$, we define
the normalization constant $C$ such that
\begin{eqnarray}
&&\mathcal{Z}=\mathcal{Z}_\mathrm{vac}^{-1}\int \rmD\rho\,\exp{\left(-\beta\, \Gamma_\omega[\psi,\psi^*]\right)},\\
&&\mathcal{Z}_\mathrm{vac}=\int \rmD\rho\,\exp{\left(-\beta E[\psi,\psi^*]\right)}.
\label{Zref}
\end{eqnarray}
Thus the partition function is normalized by means of its vacuum value $\mathcal{Z}_\mathrm{vac}$
in order to be $\mathcal{Z}(\nu=0)=1$. Our choice guarantees the elimination of purely fluctuating
degrees of freedom, which are not fixed by the chemical potential.

For a system within the ball $B$, naturally described in spherical coordinates
$(r,\Omega)$ (where $\Omega=(\theta,\varphi)$ represents a point on the unit sphere),
one has
\begin{eqnarray}
&&\Delta=\frac{1}{r}\frac{\partial^2}{\partial r^2}r+\frac{1}{r^2}\,\Delta_\Omega,\\
&&\Delta_\Omega=\frac{1}{\sin{\theta}}\frac{\partial}{\partial\theta}
\left(\sin{\theta}\frac{\partial}{\partial\theta}\right)
+\frac{1}{\sin^2{\theta}}\frac{\partial^2}{\partial\varphi^2},\\
&&\delta^3(\bfr-\bfr^\prime)=\frac{1}{r\,r^\prime\sin{\theta}}\,
\delta(r-r^\prime)\,\delta(\theta-\theta^\prime)\,
\delta(\varphi-\varphi^\prime).
\end{eqnarray}

Hereafter, we use the set of real orthonormal functions
$\{\Phi_{nlm}(r,\Omega)|n\in\mathbb{N};\ l=0,1,\dots;\ -l\leq m\leq l\}$~\cite{Greb13,Greb19}:
\begin{eqnarray}
&&\Phi_{nlm}(r,\Omega)=\sqrt{\frac{2}{R^3}}\,N_{ln}\,
j_l\left(a_{ln}\frac{r}{R}\right)\,y_{lm}(\Omega),\\
&&
N_{ln}=\frac{a_{ln}}{\sqrt{a^2_{ln}+h_l(h_l-1)-l(l+1)}\,j_l(a_{ln})},\\
&&
\int_B \Phi_{nlm}(r,\Omega)\,\Phi_{n^\prime l^\prime m^\prime}(r,\Omega)\,\rmd\bfr
=\delta_{n,n^\prime}\,\delta_{l,l^\prime}\,\delta_{m,m^\prime},\\
&&\rmd\bfr=r^2\,\rmd r\,\rmd\Omega=r^2\sin{\theta}\,\rmd r\,\rmd\theta\,\rmd\varphi,
\nonumber
\end{eqnarray}
 which is the solution to the problem:
\begin{eqnarray}
&&\Delta\,\Phi_{nlm}=-\left(\frac{a_{ln}}{R}\right)^2\,\Phi_{nlm},\\
&&\left.\left(r\partial_r\Phi_{nlm}(r,\Omega)+h_l\,\Phi_{nlm}(r,\Omega)\right)\right|_{r=R}=0,
\label{c1}
\end{eqnarray}
solved in terms of the spherical Bessel functions,
\begin{equation}
j_l(z)=z^l\left(-\frac{1}{z}\frac{\rmd}{\rmd z}\right)^l\frac{\sin{z}}{z},\qquad
y_l(z)=-z^l\left(-\frac{1}{z}\frac{\rmd}{\rmd z}\right)^l\frac{\cos{z}}{z},
\end{equation}
and the orthonormal real spherical harmonics,
\begin{equation}\nonumber
y_{lm}(\Omega)=
\begin{cases}
\sqrt{2}K^m_l\,P^m_l(\cos{\theta})\,\cos{(m\varphi)},& m>0\\
K^0_l\,P_l(\cos{\theta}),& m=0\\
\sqrt{2}K^m_l\,P^{|m|}_l(\cos{\theta})\,\sin{(|m|\varphi)},& m<0
\end{cases}
\end{equation}
\begin{equation}
K^m_l=\sqrt{\frac{2l+1}{4\pi}\frac{(l-|m|)!}{(l+|m|)!}}.
\end{equation}
Here, $P^m_l(z)$ is the Legendre polynomial.

The set of $\{a_{ln}\}$, depended on $h_l$, determines the system spectrum,
where each number $a_{ln}$ is the $n$th positive solution of equation:
\begin{equation}\label{c2}
a_{ln}j^{\prime}_l(a_{ln})+h_l\,j_l(a_{ln})=0,\quad
0<a_{l\,1}<a_{l\,2}<\ldots.
\end{equation}
Actually, using this relation at finite $h_l$, we take a possibility to express
the derivative $j^\prime_l$ through $j_l$ when the mixed boundary conditions are imposed.

Note that the conditions (\ref{c1}), (\ref{c2}) appear due to the physical necessity
of continuation of an attraction potential (including a rotational one) beyond the ball $B$.
At $r\geq R$ the potential should be of the (multipole expansion) form~\cite{Chandra33}:
\begin{eqnarray}
&&V_\mathrm{out}(\bfr)=c_{-1}+\sum\limits_{l=0}^\infty c_l\left(\frac{R}{r}\right)^{l+1}\, P_l(\cos{\theta}),
\label{Vout}\\
&&\Delta V_\mathrm{out}(\bfr)=0,\quad r\geq R,\nonumber
\end{eqnarray}
 $c_l$ are constants. Combining two boundary conditions,
\begin{eqnarray}
&&V_\mathrm{in}(\bfr)|_{r=R}=V_\mathrm{out}(\bfr)|_{r=R},\nonumber\\
&&\partial_rV_\mathrm{in}(\bfr)|_{r=R}=\partial_rV_\mathrm{out}(\bfr)|_{r=R},
\label{BC}
\end{eqnarray}
we are able to determine $h_l$ (see below).

Setting $\omega=0$, the size $R_{\kappa}=\pi/\kappa$ of bosonic cloud arises
as the result of competition between repulsion, associated with potential $U$,
and attraction, caused by gravitation.
The case of $R<R_{\kappa}$, fulfilling the condition of $\rho(\bfr)$ positivity,
reflects some geometrical constraint, imposed on $R$. However, a model with
independent $R$ looks more general.

Exploiting the completeness of $\{\Phi_{nlm}(r,\Omega)\}$, we write
\begin{eqnarray}
&&\delta^3(\bfr-\bfr^\prime)=
\sum\limits_{n,l,m} \Phi_{nlm}(r,\Omega)\,
\Phi_{nlm}(r^\prime,\Omega^\prime),\\
&&\rho(\bfr)=\frac{1}{\sqrt{V}}\sum\limits_{n,l,m} \xi_{nlm}\,
\Phi_{nlm}(r,\Omega),
\end{eqnarray}
where $\xi_{nlm}$ are new dimensionless variables.

Solution to the general problem in the ball $B$,
\begin{equation}
\frac{\Delta+p^2}{\Delta+q^2}\,G_{p,q}(\bfr|\bfr^\prime)=\delta^3(\bfr-\bfr^\prime),\qquad
\left.G_{p,q}(\bfr|\bfr^\prime)\right|_{|\bfr|=0}<\infty,
\end{equation}
can be found in the form:
\begin{eqnarray}
&&G_{p,q}(\bfr|\bfr^\prime)=\delta^3(\bfr-\bfr^\prime)+(q^2-p^2)\,g_p(\bfr|\bfr^\prime),
\label{GF1}\\
&&(\Delta+p^2)\,g_p(\bfr|\bfr^\prime)=\delta^3(\bfr-\bfr^\prime),\\
&&g_p(\bfr|\bfr^\prime)=\sum\limits_{n,l,m}
\frac{
\Phi_{nlm}(r,\Omega)\,
\Phi_{nlm}(r^\prime,\Omega^\prime)}
{p^2-(a_{ln}/R)^2}.
\end{eqnarray}

Performing summation over $\{a_{ln}\}$ accordingly to \cite{Greb19}, the Green's function
$g_p(\bfr|\bfr^\prime)$ of Helmholtz operator is presented as
\begin{eqnarray}
&&\hspace*{-2mm}
g_p(\bfr|\bfr^\prime)=\sum\limits_{l,m} g_{p\,(l)}(r|r^\prime)\,
y_{lm}(\Omega)\,y_{lm}(\Omega^\prime),\\
&&\hspace*{-2mm}
g_{p\,(l)}(r|r^\prime)=p\, j_l(p\,r_{<})\left[y_l(p\,r_{>})-j_l(p\,r_{>})\,H_l(pR)\right],\\
&&\hspace*{-2mm}
H_l(z)=\frac{zy^\prime_l(z)+h_ly_l(z)}{zj^\prime_l(z)+h_lj_l(z)},\nonumber\\
&&\hspace*{-2mm}
r_{<}=\mathrm{min}(r,r^\prime),\qquad r_{>}=\mathrm{max}(r,r^\prime).\nonumber
\end{eqnarray}

The introduced notions allow us to obtain algebraic expressions for the integrals:
\begin{eqnarray}
E[\psi,\psi^*]&=&\frac{U}{2}\int_B\rmd\bfr\int_B\rmd\bfr^\prime
\rho(\bfr)\,G_{0,\kappa}(\bfr|\bfr^\prime)\,\rho(\bfr^\prime)\nonumber\\
&=&\frac{U}{2V}\sum\limits_{n,l,m}\left[1-\left(\frac{\kappa R}{a_{ln}}\right)^2\right]
(\xi_{nlm})^2,\\
\mathcal{N}[\rho]&=&\sqrt{6}\,h_0\,\sum\limits_{n=1}^\infty\frac{\xi_{n\,0\,0}}{a_{0\,n}\sqrt{a^2_{0\,n}+h_0(h_0-1)}}.
\end{eqnarray}
The Gaussian integral calculation requires the positivity of all terms of $E$, that is
realized under condition $\kappa R<\mathrm{min}\,a_{ln}$. Since $R=R_\kappa$ within
the basic model without rotation~\cite{Harko11}, we strongly demand $\pi<\mathrm{min}\,a_{ln}$.
We will use this criterion by choosing $h_l$ (and $\{a_{ln}\}$).

One also finds
\begin{eqnarray}
\hspace*{-0.7cm}
\frac{1}{R^2}\int_B r^2\left(1-P_2(\cos{\theta})\right)\rho(\bfr)\rmd\bfr&=&
\sqrt{6}\sum\limits_{n=1}^\infty
\frac{2+h_0-6h_0/a^2_{0\,n}}{a_{0\,n}\sqrt{a^2_{0\,n}+h_0(h_0-1)}}\,\xi_{n\,0\,0}
\nonumber\\
&&
-\sqrt{\frac{6}{5}}\sum\limits_{n=1}^\infty
\frac{2+h_2}{a_{2\,n}\sqrt{a^2_{2\,n}+h_2(h_2-1)-6}}\,\xi_{n\,2\,0},
\end{eqnarray}
where (\ref{id5}), (\ref{id6}) are used.

Evaluating the Jacobian of transformation $\rho(\bfr)\mapsto\xi_{nlm}$ as
\begin{equation}
\sqrt{\mathrm{det}\left\|V\int_B\frac{\partial\rho(\bfr)}{\partial\xi_{nlm}}
\frac{\partial\rho(\bfr)}{\partial\xi_{n^\prime l^\prime m^\prime}}\,\rmd\bfr\right\|}=1,
\end{equation}
one has an equivalence of the path integration measures:
\begin{equation}
\int \rmD\rho=\prod\limits_{n,l,m}\int_{-\infty}^\infty\rmd\xi_{nlm},
\end{equation}

Integration over the variables $\{\xi_{nlm}|n=1,2,\ldots;\ l=0,1,2,\dots;\ -l\leq m\leq l\}$
results in
\begin{equation}
\mathcal{Z}=\mathcal{Z}_0\,\mathcal{Z}_2,\quad
\mathcal{Z}_l=\exp{\left(\sum\limits_{n=1}^\infty F_{nl}\right)},\quad l=0,2;
\end{equation}
while $\mathcal{Z}_\mathrm{vac}\sim D^{-1/2}(x,q)$, where
\begin{equation}
D(x,q)=\prod\limits_{n=1}^\infty\prod\limits_{l=0}^\infty
\left[q\left(1-\frac{x^2}{a^2_{ln}}\right)\right]^{2l+1}
\end{equation}
is the Fredholm determinant $\mathrm{Det}\left\|q(1+\kappa^2\Delta^{-1})\right\|$
for operator, defined in the ball $B$; $x=\kappa R$ and $q=\beta U/V$.

The partition functions $\mathcal{Z}_0$ and $\mathcal{Z}_2$ are determined by
\begin{eqnarray}
&&\hspace{-4mm}
F_{n0}=\frac{V\beta/(3U)}{a^2_{0\,n}-x^2}
\frac{\left[3h_0\mu+m\omega^2R^2\left(2+h_0-6h_0/a^2_{0\,n}\right)\right]^2}{a^2_{0\,n}+h_0(h_0-1)},\\
&&\hspace{-4mm}
F_{n2}=\frac{V\beta/(15U)}{a^2_{2\,n}-x^2}
\frac{m^2\omega^4R^4(h_2+2)^2}{a^2_{2\,n}+h_2(h_2-1)-6}.
\end{eqnarray}
We immediately see that the functions $F_{nl}$ inherit a singularity property
with respect to parameter $x=\kappa R$.

Further, let us define the grand thermodynamic potential
$\Omega_\omega(T,V,\mu)=-T\sum_{n=1}^\infty (F_{n0}+F_{n2})$ such that
\begin{eqnarray}
-\frac{\Omega_\omega}{V}&=&\frac{\mu^2}{2U}\,A_1(x,h_0)+\mu vA_2(x,h_0)
+U\frac{v^2}{2}\,[A_3(x,h_0)+B(x,h_2)];\\
v&=&\frac{\omega^2}{2\pi Gm}.
\end{eqnarray}
Performing a formal summation over $\{a_{ln}\}$ by the rules from \cite{Greb19},
the functions $A_i(x,h_0)$ and $B(x,h_2)$ become
\begin{eqnarray}
A_1(x,h_0)&=&\frac{3h_0j_1(x)}{x[xj^\prime_0(x)+h_0j_0(x)]},\\
A_2(x,h_0)&=&1+\frac{x^2(2+h_0)-6h_0}{2x[xj^\prime_0(x)+h_0j_0(x)]}\,j_1(x),\\
A_3(x,h_0)&=&\frac{[x^2(2+h_0)-6h_0]^2}{12h_0x[xj^\prime_0(x)+h_0j_0(x)]}\,j_1(x)+x^2\frac{4h_0+5}{15h_0}-1,\\
B(x,h_2)&=&\frac{2+h_2}{60}\,x^2\left[\frac{(2+h_2)\,j_2(x)}{x j^\prime_2(x)+h_2j_2(x)}-1\right].
\end{eqnarray}
In practice, the auxiliary formulas (\ref{id7})--(\ref{id9}) are used.

Using the potential $\Omega_\omega(T,V,\mu)$, the mean number of particles $\mathcal{N}=V\bar\rho$,
mean energy $\mathcal{E}$ and the pressure $P$ are found
\begin{eqnarray}
\bar\rho&=&\frac{\mu}{U}\,A_1+v A_2\label{N0},\\
\frac{\mathcal{E}}{V}&=&\frac{\mu^2}{2U}\,A_1-U\frac{v^2}{2}\,(A_3+B),
\label{Er}\\
P&=&\frac{\mu^2}{2U}\,A_1+\mu vA_2+U\frac{v^2}{2}\,(A_3+B)\nonumber\\
&=&\frac{U}{2A_1}\left[\bar\rho^2+v^2\left(A_1A_3-A^2_2+A_1B\right)\right].
\label{Pr}
\end{eqnarray}
The total entropy, $S\equiv-\partial_T\Omega_\omega=0$, vanishes here.
We see that the ellipsoidal form of the cloud results in contributions to
$\mathcal{E}$ and $P$, represented by $B(x,h_2)$.

In the absence of rotation ($\omega=v=0$), the sign of these functions is
mainly determined by $A_1$ at $x=\pi$:
\begin{eqnarray}
&&A_1(\pi,h_0)=-\frac{3h_0}{\pi^2},\quad A_2(\pi,h_0)=\frac{3h_0}{\pi^2}-\frac{h_0}{2},\nonumber\\
&&A_3(\pi,h_0)=-\frac{3h_0}{\pi^2}+1+h_0-\pi^2\left(\frac{h_0}{12}+\frac{1}{15}\right),\nonumber\\
&&B(\pi,h_2)=\frac{2+h_2}{60}\frac{15\pi^2-\pi^4}{\pi^2-9+3h_2}.
\end{eqnarray}

To obey $\mathcal{E}>0$ and $P>0$ (the sign of $\mathcal{N}$ depends also on $\mathrm{sign}(\mu)$),
we should require $h_0<0$. Indeed, putting $h_0=-1$, we find that $a_{0\,1}\approx 4.27479>\pi$.
At the same time, $a_{0\,1}=\pi/2$ for $h_0=1$ that leads to the divergence of Gaussian integral
$\mathcal{Z}$, and negative $\mathcal{E}<0$, $P<0$. Note that the factor $A_1=3/\pi^2$
is already appeared in \cite{Harko11} on the base of differential calculus. Let us announce
immediately that $h_2=3$ to give $a_{2\,1}\approx 4.4934$. This is shown in \cite{Harko18,Chandra33}
and is reproduced below.

Thus the chemical potential turns out to be $\mu>0$, and the central density of non-rotating
cloud equals to $\mu/U$. To find $\mu$ at $T=0$ and $x=\pi$, we invert
the formula for $\mathcal{N}$. One has
\begin{equation}\label{mu}
\mu=G\frac{m^2\mathcal{N}}{R_{\kappa}}+Uv\left(\frac{\pi^2}{6}-1\right),\quad
R_\kappa=\frac{\pi}{\kappa}.
\end{equation}
This is a work of attraction forces by transferring particle from infinitely distanced
region to the cloud boundary.

On the other hand, extremizing the functional $\Gamma_\omega[\psi,\psi^*]$
with the use of the Poisson brackets~(\ref{PB1}),
\begin{equation}
\{\psi(\bfr),\Gamma_\omega[\psi,\psi^*]\}=0,
\end{equation}
one obtains the stationary Gross--Pitaevskii equation for finding
a condensate density $\rho_\omega$ at $T=0$:
\begin{equation}\label{eq1}
U\left(1+\kappa^2\Delta^{-1}\right)\rho_\omega=\mu+\frac{m\omega^2r^2}{3}\left(1-P_2(\cos{\theta})\right).
\end{equation}

Its solution is given by the integral:
\begin{eqnarray}
\rho_\omega(r,\theta)&=&U^{-1}\int_B G_{\kappa,0}(\bfr|\bfr^\prime)\,\nu(\bfr^\prime)\,\rmd\bfr^\prime
\nonumber\\
&=&v+(\rho_c-v)j_0(\kappa r)-
\frac{v}{6}\frac{x^2\,(2+h_2)}{xj^\prime_2(x)+h_2j_2(x)}j_2(\kappa r)P_2(\cos{\theta});\\
\rho_c&=&v+\frac{h_0\left[\mu/U-v+v\, x^2\,(2+h_0)/(6h_0)\right]}{xj^\prime_0(x)+h_0j_0(x)},
\label{rhoc}
\end{eqnarray}
where the central density, $\rho_c=\rho_\omega(0,\theta)$, depends on $x$. 

Switching off the gravity, $x=\kappa=0$, and rotation, $v=\omega=0$, one has a homogeneous
distribution $\rho_0(r)=(\mu/U)\,\Theta(R-r)$, which is limited by the boundary and expressed
in terms of the Heaviside function $\Theta(z)$.

Integrating $\rho_\omega(r,\theta)$ over the ball $B$, we obtain the number of particles $\mathcal{N}[\rho_\omega]$
which coincides with~(\ref{N0}). Moreover, the correspondence between the partition function
and the solution to Gross--Pitaevskii equation consists in the relation:
\begin{equation}
-2\Omega_\omega=\int_B\nu(\bfr)\rho_\omega(r,\theta)\rmd\bfr.
\end{equation}

Following (\ref{GF1}), we find the effective potential:
\begin{eqnarray}
V_\mathrm{in}(r,\theta)&=&U^{-1}\int_B g_{\kappa}(\bfr|\bfr^\prime)\,\nu(\bfr^\prime)\,\rmd\bfr^\prime
\nonumber\\
&=&-\frac{v}{\kappa^2}+\frac{\mu}{\kappa^2U}\left[1-\frac{h_0j_0(\kappa r)}{xj^\prime_0(x)+h_0j_0(x)}\right]
-\frac{v}{6\kappa^2}\frac{x^2(2+h_0)-6h_0}{xj^\prime_0(x)+h_0j_0(x)}\,j_0(\kappa r)
\nonumber\\
&&+v\frac{r^2}{6}\left[1-P_2(\cos{\theta})\right]
+\frac{v}{6\kappa^2}\,\frac{x^2\,(2+h_2)}{xj^\prime_2(x)+h_2j_2(x)}\,j_2(\kappa r)\,P_2(\cos{\theta}),
\end{eqnarray}
which satisfies the equation inside the ball $B$:
\begin{equation}
\Delta V_\mathrm{in}=\rho_\omega,\quad r\leq R.
\end{equation}
We also note the relation between $V_\mathrm{in}$, $\rho_\omega$
and the chemical potential $\nu(\bfr)$:
\begin{equation}
\rho_\omega(r,\theta)=U^{-1}\,\nu(\bfr)-\kappa^2\,V_\mathrm{in}(r,\theta).
\end{equation}

To determine parameters $h_0$ and $h_2$, we connect the potentials
$V_\mathrm{in}$ and $V_\mathrm{out}$ from (\ref{Vout}) by means of the boundary conditions
(\ref{BC}). It is clear that $c_{-1}$, $c_0$ and $c_2$ are supposed to be non-vanishing
while the others are dropped out. 

Since the potentials relation should hold for any admissible $x$, we fix $x=\pi$ for
the sake of simplicity. Using the central density $\rho_c$ as independent variable
instead of chemical potential $\mu$, we arrive at
\begin{eqnarray}
V_\mathrm{in}(r,\theta)&=&-\frac{\rho_c}{\kappa^2}\left[\frac{1}{h_0}+j_0(\kappa r)\right]
+v\frac{r^2}{6}[1-P_2(\cos{\theta})]+\nonumber\\
&&+\frac{v}{\kappa^2}\,j_0(\kappa r)-\frac{v}{\kappa^2}\frac{2+h_0}{6h_0}\left(\pi^2-\frac{6h_0}{2+h_0}\right)+
\nonumber\\
&&+\frac{v}{\kappa^2}\frac{\pi^4}{6}\frac{2+h_2}{\pi^2-9+3h_2}\,j_2(\kappa r)\, P_2(\cos{\theta}).
\end{eqnarray}

Let us first relate the terms with $P_2(\cos{\theta})$. We obtain a pair of equations:
\begin{eqnarray}
c_2&=&-\frac{v}{\kappa^2}\frac{\pi^2}{6}\left\{1-\frac{6+3h_2}{\pi^2-9+3h_2}\right\},
\nonumber\\
-3c_2&=&-\frac{v}{\kappa^2}\frac{\pi^2}{6}\left\{2-\frac{(2+h_2)(\pi^2-9)}{\pi^2-9+3h_2}\right\}.
\end{eqnarray}
Solving, we derive that $h_2=3$ as it must be \cite{Chandra33}.

Substituting $h_2=3$ and $h_0=1$ into $V_\mathrm{in}$, we reproduce (up to multiplier $4\pi G$)
the effective potential from \cite{Harko18}. It results also in $c_{-1}=0$.

However, we have already noted above that the substitution $h_0=1$ leads to the invalid Gauss
integral $\mathcal{Z}$ and the negative thermodynamic functions like $\mathcal{E}$ and $P$.
A possibility of replacement of $h_0=1$ with $h_0=-1$ can be justified by uncertainty of
three coefficients $c_{-1}$, $c_0$ and $h_0$ under two (boundary) conditions imposed on
the terms with $r^{-1}$ and $r^0$.

Performing the physical analysis of the solution, we set $h_2=3$ and $h_0=-1$.
It immediately limits the value of the parameter $x<a_{0\,1}\approx4.275$.

Averaging distribution $\rho_\omega(r,\theta)$ over the angles,
\begin{equation}\label{rhoav}
\tilde\rho_\omega(r)=\langle\rho_\omega(r,\theta)\rangle=\rho_c[\eta+(1-\eta)\,j_0(\kappa r)],
\end{equation}
we find the mass profile $M(r)$ and the tangential velocity $\upsilon_\mathrm{tan}(r)$ of a probe particle:
\begin{eqnarray}
M(r)&=&4\pi m\,\int_0^r\tilde\rho_\omega(r_*)\,r^2_*\rmd r_*
\nonumber\\
&=&\frac{4\pi m\,\rho_c r^3}{3}\left[1+(1-\eta)\,\frac{3}{\kappa r}\,j_1(\kappa r)\right],\\
\upsilon_\mathrm{tan}(r)&=&\sqrt{G\,\frac{M(r)}{r}}
\nonumber\\
&=&\upsilon_0\,\sqrt{\frac{\eta}{3}\,(\kappa r)^2+(1-\eta)\,(\kappa r)\,j_1(\kappa r)},
\label{vtan}
\end{eqnarray}
where $\upsilon_0=\sqrt{U\rho_c/m}$ and $\eta=v/\rho_c$.

\begin{figure}
\begin{center}
\includegraphics[width=0.6\linewidth]{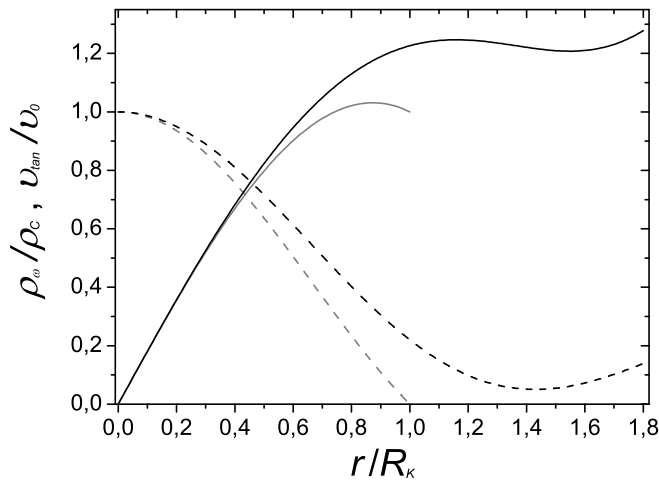}
\end{center}
\vspace*{-5mm}%\vspace*{8mm}
\caption{The averaged particle distribution (\ref{rhoav}) (dashed lines) and tangential velocity
(\ref{vtan}) (solid lines)  as the dimensionless functions of radial distance $r/R_\kappa$ at
different $\eta$. Gray lines correspond to $\eta=0$ (without rotation), while black ones are
for $\eta=0.22$.}
\end{figure}

To fix radius $R$, we impose the condition $\tilde\rho_\omega(R)=0$ which immediately gives us
the relation between $0\leq\eta<1$ and $\pi\leq x<a_{0\,1}$:
\begin{equation}\label{etax}
\eta=-\frac{j_0(x)}{1-j_0(x)}.
\end{equation}
It allows us to extract admissible values of $v$ and $R$ (at given $\kappa$ and $\rho_c$)
and resolves the problem of applicability of the Thomas--Fermi approximation.

The rotational curves $\upsilon_\mathrm{tan}(r)$ at $0\leq\eta<1$ are successfully
applied to explain observational data of dwarf galaxies~\cite{Harko11,Harko18}. On the other
hand, the particle distribution at $\eta>\eta_{\mathrm{max}}$ ($\eta_{\mathrm{max}}\simeq0.1785$
over the interval $x\in[\pi;2\pi]$) has no the limiting radius $R$ (see Fig.~1). Its absence
complicates the interpretation of the results in the Thomas--Fermi approximation. In practice,
the restrictions made in \cite{KKG} are used.

Since the boson cloud has the ellipsoidal shape revealed by equation
$\rho_\omega(r,\theta)=\mathrm{const}$, some macroscopic characteristics can be corrected due to
this fact~\cite{Harko18,Chandra33}. Here we omit these considerations.

%%%%%%%%%%%%%%%%%%%%%%%%%%%%%%%%%%%%%%%%%%%%%%%%%%%%%%%%%%%%%%%%%%%%%%%%%%%%%%%%%%%%%%%%%%%%%
\section{The Modified Gross--Pitaevskii Equation}

Now we would like to show how additional interaction can be brought into the system using
the partition function formalism. Here we restrict ourselves to the inclusion of a potential
which simulates the rigid rotation studied before. Technically, we change the commutation
relations for the macroscopic wave function to obtain a nontrivial measure of path integration,
which after simple manipulations can be regarded as a correction to energy. In a geometric
sense, we consider a curved functional space.

Thus we postulate the deformed Poisson brackets:
\begin{eqnarray}
&&\{\psi(\mathbf{r}_1),\psi(\mathbf{r}_2)\}=\{\psi^*(\mathbf{r}_1),\psi^*(\mathbf{r}_2)\}=0,
\nonumber\\
&&\{\psi(\mathbf{r}_1),\psi^*(\mathbf{r}_2)\}=-\rmi F(\bfr_1;\rho)\,\delta^3(\mathbf{r}_1-\mathbf{r}_2),
\label{def}
\end{eqnarray}
where $F(\bfr_1;\rho)$ is a functional of density $\rho=|\psi|^2$ at the point $\bfr_1$
(or $\bfr_2$).

Using again the Madelung transformation (\ref{Mad}), the relation (\ref{def}) means
\begin{equation}\label{rfF}
\{\rho(\bfr_1),\phi(\bfr_2)\}=F(\bfr_2;\rho)\,\delta^3(\bfr_1-\bfr_2),
\end{equation}
that leads to redefinition $\phi(\bfr)=\tilde\phi(\bfr)\,F(\bfr;\rho)$, where $\tilde\phi(\bfr)$
is a canonical variable, which satisfies (\ref{rfF}) with $F=1$.

Constructing a partition function, we take (\ref{def}) into account and define
\begin{eqnarray}
&&\mathcal{Z}_*=C\int \rmD\psi\,\rmD\psi^*\,J[\psi,\psi^*]\exp{\left(-\beta \Gamma_0[\psi,\psi^*]\right)},
\label{Zsec1}\\
&&\Gamma_0[\psi,\psi^*]=E[\psi,\psi^*]-\mu\mathcal{N}[\rho],\\
&&J[\psi,\psi^*]=\prod_{\bfr\in B}F(\bfr;\rho),
\end{eqnarray}
where $\mathcal{N}$ and $E$ are given by (\ref{Num}) and (\ref{EnFn}),
respectively.

Applying the relation $\det{\hat A}=\exp{(\Tr\ln{\hat A})}$ to
diagonal matrix ${\hat A}=\left\|\{\psi(\mathbf{r}_1),\psi^*(\mathbf{r}_2)\}\right\|$,
we obtain the temperature dependent contribution to energy:
\begin{equation}
E_*[\psi,\psi^*]=E[\psi,\psi^*]-\frac{T}{V}\int_B\ln{F(\bfr;\rho)}\,\rmd\bfr.
\end{equation}

Since this functional depends on $\rho$ (but not $\phi$), we formulate our problem
by specifying the normalization as (see (\ref{Zref}))
\begin{eqnarray}
&&\mathcal{Z}_*=\mathcal{Z}_\mathrm{vac}^{-1}\int \rmD\rho\,\exp{\left(-\beta\, \Gamma_*[\psi,\psi^*]\right)},\\
&&\Gamma_*[\psi,\psi^*]\equiv E_*[\psi,\psi^*]-\mu\mathcal{N}[\rho].
\end{eqnarray}

Extremizing $\Gamma_*[\psi,\psi^*]$, $\delta\Gamma_*[\psi,\psi^*]/\delta\rho(\bfr)=0$,
we arrive at the modified (static) Gross--Pitaevskii equation:
\begin{eqnarray}
&&U\left(1+\kappa^2\Delta^{-1}\right)\rho(\bfr)=\mu+T\,s(\bfr),\label{eqT1}\\
&&s(\bfr)=\frac{1}{V}\int_B\frac{1}{F(\bfr_1;\rho)}\frac{\delta F(\bfr_1;\rho)}{\delta\rho(\bfr)}\,\rmd\bfr_1.
\end{eqnarray}

Despite the uncertainty of the boundary radius of the slowly rotating condensate
dark matter, the obtained distribution, as we see in \cite{Harko18,KKG}, allows us
to successfully describe the rotation curves of the halo of dwarf galaxies. This
prompts us in our heuristic approach to assume that $s(\bfr)\sim(r/R)^2$. Below we
discuss the conditions imposed on $F$ and the explicit, albeit hypothetical, form
of function $s(\bfr)$. 

As argued above, we consider the case
\begin{equation}
s(\bfr)=\tmu\left\langle\frac{3}{2}\frac{\bfr^2}{R^2}\sin^2{\theta}-1\right\rangle
=\tmu\left(\frac{r^2}{R^2}-1\right),
\label{S1}
\end{equation}
where $0\leq\tmu$ is the model parameter, regarded as the strength
of interaction; $\langle(\ldots)\rangle$ means the averaging over the angles.

Now it seems important to evaluate temperature $T$, which is substituted into (\ref{eqT1})
and replaces the expression $\sim m\omega^2R^2$ in the model with rotation. Using
the characteristic scales of the BEC dark matter from \cite{KKG}, one finds
\begin{eqnarray}
\tmu T&=&\frac{m\omega^2R^2}{k_\mathrm{B}}
\nonumber\\
&=&T_0\left(\frac{mc^2}{10^{-17}\,\mathrm{eV}}\right)
\left(\frac{\omega}{10^{-16}\,\mathrm{s}^{-1}}\right)^2\left(\frac{R}{1\,\mathrm{kpc}}\right)^2,\\
T_0&\simeq& 1.23\cdot10^{-23}\,\mathrm{K},\qquad
k_\mathrm{B}T_0\simeq1.06\cdot10^{-27}\,\mathrm{eV},
\end{eqnarray}
where we rewrote all quantities in appropriate units;
$k_\mathrm{B}\simeq8.6173\cdot10^{-5}\,\mathrm{eV}\cdot\mathrm{K}^{-1}$ is the Boltzmann constant.

Thus this value is small in the absolute sense and several tens of orders of magnitude
lower than the critical temperature of the non-interacting Bose condensate~\cite{HLLM15}.

The most significant argument for us, when the form of function $s(\bfr)$ is determined,
is the property of the generating functional: $F\to0$ at $\mathcal{N}\to\infty$, that is,
it leads to the commuting macroscopic wave functions $\psi(\bfr)$ and $\psi^*(\bfr)$.
This is similar to replacing the quantum creation and annihilation operators in the
coordinate representation with the commuting complex numbers.

Thus the simplest form of $F$ is
\begin{eqnarray}
&&F(\bfr;\rho)=\exp{\left[-\lambda^3\, \rho(\bfr)\left(1-\frac{\bfr^2}{R^2}\right)\right]},
\quad r\leq R,\label{FF}\\
&&\int_B\ln{F(\bfr;\rho)}\,\frac{\rmd\bfr}{V}=
-\tmu\left(\mathcal{N}[\rho]-\frac{\sigma^2[\rho]}{R^2}\right),\\
&&\sigma^2[\rho]=\int_B\bfr^2\rho(\bfr)\,\rmd\bfr,\quad \tmu=\frac{\lambda^3}{V},
\nonumber
\end{eqnarray}
where $\lambda$ is the correlation length assumed to be larger than
interparticle distance, $\lambda^3\gg V/\mathcal{N}$, for the sake
of the BEC existence. In the case of ideal boson gas the thermal
wavelength $\lambda_T=\sqrt{2\pi\hbar^2/(mk_\mathrm{B}T)}$ can be used.
Hereafter we suppose $\lambda$ is independent of thermodynamic variables.

Therefore the solution to (\ref{eqT1}) is
\begin{equation}\label{solT}
\rho_T(r)=\tilde{v}+(\tilde\rho_c-\tilde{v})\,j_0(\kappa r),\qquad
\tilde{v}=\frac{6\tmu}{x^2}\frac{T}{U},
\end{equation}
where the central density $\tilde\rho_c$ is determined from (\ref{rhoc}) by replacing
$\mu$ and $v$ with $\mu-\tmu\,T$ and $\tilde{v}$, respectively.

Introducing $\eta=\tilde{v}/\tilde\rho_c$ and appealing to (\ref{etax}), we conclude that
the admissible radius $R=x/\kappa$ is implicit function of temperature $T$;
$R=\pi/\kappa$ at $T=0$. 

As it was expected, (\ref{solT}) coincides formally with (\ref{rhoav}) and leads to outcomes
similar for the rotation curves~(\ref{vtan}). However, thermodynamics of such a system looks
rather different.

Indeed, using the results of previous Section, when
\begin{equation}
\frac{\sigma^2[\rho]}{R^2}=\sqrt{6}\sum\limits_{n=1}^\infty
\frac{2+h_0-6h_0/a^2_{0\,n}}{a_{0\,n}\sqrt{a^2_{0\,n}+h_0(h_0-1)}}\,\xi_{n\,0\,0},
\end{equation}
the partition function for this model becomes
\begin{eqnarray}
\mathcal{Z}_*&=&\exp{(-\beta \Omega_{\tmu})},\\
-\frac{\Omega_{\tmu}}{V}&=&\frac{(\mu-\tmu T)^2}{2U}\,A_1(x,h_0)+(\mu-\tmu T)\tilde{v}A_2(x,h_0)
+U\frac{\tilde{v}^2}{2}A_3(x,h_0),
\end{eqnarray}
where $\Omega_{\tmu}(T,V,\mu)$ is the grand potential. It is assumed again that $h_0=-1$.

One can verify that
\begin{equation}
-2\Omega_{\tmu}=\int_B\left(\mu+Ts(\bfr)\right)\rho_T(r)\,\rmd\bfr.
\end{equation}

The mean number of particles $\mathcal{N}=V\bar\rho_T$, mean energy $\mathcal{E}$
and the pressure $P$ are determined by
\begin{eqnarray}
\bar\rho_T&=&\frac{\mu-\tmu T}{U}\,A_1+\tilde{v} A_2\label{N0T},\\
P&=&\frac{(\mu-\tmu T)^2}{2U}\,A_1+(\mu-\tmu T)\tilde{v}A_2+U\frac{\tilde{v}^2}{2}A_3
\nonumber\\
&=&\frac{U}{2A_1}\left[\bar\rho^2_T+\tilde{v}^2(A_1A_3-A^2_2)\right],
\label{PT}\\
\mathcal{E}&=&PV.
\label{ET}
\end{eqnarray}
Interestingly, (\ref{ET}) coincides with the same thermodynamic relation for
the non-rotating BEC at $T=0$ (compare (\ref{Er}) and (\ref{Pr}) at $v=0$) \cite{Harko11}.
Physically, this can be explained as follows: In the case of deformations, we consider
the original system, but of unusual particles. This is confirmed by models of an ideal
gas consisting of bosons with deformed commutation relations~\cite{RKG}.

In contrast to the model with rotation, the entropy does not vanish here:
\begin{eqnarray}
S&=&\int_Bs(\bfr)\rho_T(r)\,\rmd\bfr\\
&=&\frac{\lambda^3}{A_1}\left[-\bar\rho_T\left(A_1-\frac{6}{x^2}\,A_2\right)
+\tilde{v}\frac{6}{x^2}(A_1A_3-A^2_2)\right].
\end{eqnarray}
It is locally determined by the function $s(\bfr)$ such that its minimal value
corresponds to $r=0$, while it reaches a maximum at the boundary $r=R$.

Analyzing
\begin{eqnarray}
\overline{\ln{F}}&=&\int_B\ln{F(\bfr;\rho)}\,\frac{\rmd\bfr}{V}
\nonumber\\
&=&-\lambda^3\tilde\rho_c\left[\frac{2}{5}\eta+\frac{6}{x^2}(1-\eta) j_2(x)\right],\quad
\eta=\tilde{v}/\tilde\rho_c,
\end{eqnarray}
we see that $\overline{\ln{F}}<0$ for $0\leq\eta<1$ and $x<a_{0\,1}$.

Note that the non-zero temperature in our model does not lead to the
appearance of a thermal cloud (of dark matter particles and/or excitations), describing
which we can determine the critical parameters of the transition to the condensate state
(see \cite{HarMad11}). Non-condensate degrees of freedom were not included in
the partition function here. We show how the statistical properties of the condensate
of deformed bosons vary with temperature (much lower than the condensation temperature):
It enhances the strength of supplementary (local) interaction between particles, which is
inherited from deformation. This results in inhomogeneous entropy but the particle distribution
as in the case of the slowly rotating condensate of ordinary bosons (replacing $\tilde{v}$
with $v=\omega^2/2\pi Gm$).

%%%%%%%%%%%%%%%%%%%%%%%%%%%%%%%%%%%%%%%%%%%%%%%%%%%%%%%%%%%%%%%%%%%%%%%%%%%%%%%%%%%%%%%%%%%%%
\section{Discussion}

Since the observed characteristics of DM are mainly local quantities in space,
differential calculus is naturally used to find them. Within the framework of the BEC concept,
when the particles under the certain conditions are in the ground state and their number is large,
the condensate wave function in the coordinate representation is not a quantum operator, but
becomes an ordinary function obeying the Gross--Pitaevskii equation. Moreover,
the wave function squared does not describe the probability density of single particle, but rather
the spatial distribution of particles. The applicability of this approach to DM was analyzed in
\cite{Harko11,BH07} within the Thomas--Fermi approximation, when kinetic energy can be neglected
in comparison with the interaction. From the very beginning, it was assumed that particles undergo
repulsive scattering, gravitational attraction, and the whole system can rotate slowly~\cite{BH07}.
An analytical solution to the most general and static problem is obtained in \cite{Harko18}, and
its application for describing the DM characteristics is demonstrated. An analysis of the obtained
particle distribution and an attempt to fit the observed data are also undertaken in \cite{KKG}.
The very optimistic consequences, as well as the lack of thermodynamic functions in these works,
stimulated us to study this model in another approach based on calculation of the partition
function. 

Despite the fact that the partition function leads to averaged characteristics and
thermodynamic functions, we examine the Green's functions, which are a tool for finding local
quantities. This allows us to reproduce the known results~\cite{Harko18} and to make
a number of generalizations. First of all, we do not constrain $\kappa$ and $R$ by the relation
$\kappa R=\pi$ and use $x=\kappa R$ as a free parameter limited by the condition
$x<a_{0\,1}\approx4.275$. Such a requirement is a consequence of our attempt to formulate in
closed form (\ref{c2}) the boundary conditions imposed in \cite{Harko18,Chandra33}.
Varying $x$ allows one to correctly calculate partition function as the Gaussian integral
and thermodynamic functions. Adjusting $x$ in accordance with the cyclic frequency $\omega$,
we obtain the boundary radius $R$ for a model with rotation (see (\ref{etax})).

One of the academic goals of the paper is to investigate the ability to describe an interaction
by deforming commutation relations for the condensate wave function. We choose rotation
as such an interaction and show that the proposed deformation (\ref{def}), (\ref{FF}) leads to
a change in the integration measure in the expression for partition function. Formally, one gets
an effective energy functional that depends on temperature. Its variation with respect to the
wave function gives us a modified Gross--Pitaevskii equation. Indeed, this approach allows
one to derive a particle distribution similar to the model with rotation, although it becomes
dependent on temperature. Physically, it is interesting that this type of deformation leads
to entropy which is inhomogeneous in space. Guided by the commutativity condition (in the terms of
Poisson brackets) for the wave function and its complex conjugate at high particle density,
we provide a minimum of entropy at the center of the system and a maximum at the boundary.
Moreover, such a deformation does not violate the ratio between the total energy and pressure,
which occurs at zero temperature~\cite{Harko11}.

Note that, in order to reproduce the rigid rotation, the deformation functional $F$ can be
specified in different ways. Here, we want it to be given locally. However, for particle
systems much smaller than galaxies, one could use
\begin{equation}
F=\exp{\left[-\tmu\left(\mathcal{N}[\rho]-\frac{\sigma^2[\rho]}{R^2}\right)\right]}.
\end{equation}
As a result, we would get $s(\bfr)$ in the same form of (\ref{S1}).

Although in our examples the parameter $\tmu$ is assumed to be small, we did not use
any approximations. Of course, even more possibilities appear if expansion over $\tmu$ 
is allowed. It might be convenient for some purposes to introduce an alternative expression,
\begin{equation}
F=\frac{1+(1-\alpha)\tmu\,\sigma^2[\rho]/R^2}{1+\tmu\,\mathcal{N}[\rho]-\alpha\tmu\,\sigma^2[\rho]/R^2},
\end{equation}
where $0\leq\alpha\leq1$. If $\sigma^2[\rho]/R^2\to0$, we can get a system similar to Ref.~\cite{RKG},
where $\tmu$-calculus is developed.

Neglecting the contribution of kinetic energy in the Thomas--Fermi approximation (which is
justified by the estimates made~\cite{Harko19}), in the model with deformation we can expect
however the appearance of extra terms of interaction originating from it. A detailed study
of such a generalization of the model might be the subject of future research, closely
related with the concept of self-interacting DM~\cite{HMKT15,KSKS15,KKPY17,RME18}.

As suggested by the observational data, we cannot reject the possibility that
DM is self-interacting. However, the nature of DM particles is far from complete understanding.
Special attention is paid to such (theoretical) candidates as weakly interacting massive particles
(WIMPs)~\cite{A18}, axions~\cite{CM16,SNE16}, and sterile neutrinos~\cite{SN19}. We would like
also to note primordial black holes~\cite{Bel14,CKS16}, strongly interacting massive particles and superweakly
interacting particles, and even electromagnetically interacting massive particles~\cite{GKV19}
which are possibly hidden in neutral atom-like states.

Considering that among the candidates listed above there is a large number of fermions,
the formation of boson-like composites from them can play a significant role in understanding
DM, as well as, in particular, in applying the model presented here. From a descriptive point
of view, the deformation approach used can also be applied to study structured particles~\cite{GM15}.

\section*{Acknowledgments}

A.N. thanks A.M.~Gavrilik (BITP) for valuable suggestions and is indebted for partial support of
this work by the Division of physics and astronomy of NAS of Ukraine (Project No. 0117U000240).

\appendix

\section{Used Formulas}

Here we use auxiliary formulas, based on Refs.~\cite{Greb19,AS}, with $a_{ln}$ such that
$a_{ln}j^{\prime}_l(a_{ln})+h_l\,j_l(a_{ln})=0$:
\begin{eqnarray}
&&\hspace*{-3mm}
\int_0^1j_0(a_{0\,n}t)\,t^4\,\rmd t=\left[\frac{h_0+2}{a^2_{0\,n}}-\frac{6h_0}{a^4_{0\,n}}\right] j_0(a_{0\,n}),
\label{id5}\\
&&\hspace*{-3mm}
\int_0^1j_2(a_{2\,n}t)\,t^4\,\rmd t=\frac{h_2+2}{a^2_{2\,n}}\,j_2(a_{2\,n}),
\label{id6}\\
&&\hspace*{-3mm}
2\sum\limits_{n=1}^\infty\frac{1}{a^2_{ln}-x^2}=\frac{x^2-l(l+1-h_l)}{h_l\,x^2}-
\frac{x^2+h_l(h_l-1)-l(l+1)}{h_lx[x j^\prime_l(x)+h_lj_l(x)]}\,j^\prime_l(x),
\label{id7}\\
&&\hspace*{-3mm}
2\sum\limits_{n=1}^\infty\frac{1}{a^2_{ln}+h_l(h_l-1)-l(l+1)}=\frac{1}{l+h_l},
\label{id8}\\
&&\hspace*{-3mm}
2\sum\limits_{n=1}^\infty\frac{1}{a^2_{0\,n}}=\frac{2+h_0}{3h_0},\qquad
2\sum\limits_{n=1}^\infty\frac{1}{a^2_{2\,n}}=\frac{4+h_2}{7(2+h_2)}.
\label{id9}
\end{eqnarray}

%\begin{thebibliography}{000} %for 3 digits
%\begin{thebibliography}{00}  %for 2 digits

\end{document}